\documentclass[prd,twocolumn,floats,superscriptaddress,eqsecnum,nofootinbib,floatfix]{revtex4}

\usepackage{amsmath}
\usepackage{graphics, graphicx}
\usepackage{epsfig}
\usepackage{ulem}
\usepackage{color}

\def\ah{\hat a}

\def\phih{\hat\phi}

\def\1p{{(1p)}}
\def\be{\begin{equation}}
\def\ee{\end{equation}}
\def\beq{\begin{eqnarray}}
\def\eeq{\end{eqnarray}}

\def\p0{\phi_0}
\def\z0{\zeta_0}

\def\cV{{\cal V}}
\def\epa{{(0)}}
\def\epb{{(2)}}

\def\cob{}

\def\crc{}
\def\crd{}
\def\cre{}
\def\crf{}
\def\crg{}
\def\crh{}
\def\crj{}
\def\clr{}
\def\cls{}
\def\clt{}
\def\cred{}

\newcommand{\ttle}[1]{{\it #1}}

\begin{document}

\title{Arrows of time in the bouncing universes \\ of  the no-boundary quantum state}

\author{James  Hartle}
\affiliation{Department of Physics, University of California, Santa Barbara,  93106, USA}
\author{Thomas Hertog}
\affiliation{Institute for Theoretical Physics, KU Leuven, 3001 Leuven, Belgium {\it and}\\
International Solvay Institutes, Boulevard du Triomphe, ULB, 1050 Brussels, Belgium}
\bibliographystyle{unsrt}

\begin{abstract}
We  derive the arrows of time of our universe that follow from the no-boundary theory of its  quantum state (NBWF) {\crh in a minisuperspace model}. Arrows of time are viewed four-dimensionally as properties of the four-dimensional Lorentzian histories of the universe. Probabilities for these histories are predicted by the NBWF. For histories with a regular  `bounce' at a minimum radius  fluctuations are small at the bounce and grow in the direction of expansion on either side. 
 For recollapsing classical histories with big bang and big crunch singularities the fluctuations are small near one singularity and grow through the expansion and recontraction to the other  singularity. The arrow of time defined by the growth in fluctuations thus points in one direction over the whole of a recollapsing spacetime but is bidirectional in a bouncing spacetime. We argue that the electromagnetic, thermodynamic, and psychological arrows of time are aligned with the fluctuation arrow. The implications of a bidirectional arrow of time for causality are discussed. 
\end{abstract}



\maketitle

\section{Introduction}
\label{intro}

In 1897 Boltzmann wrote: ``The second law of thermodynamics can be proved from the
[time-reversible] mechanical theory, if one assumes that the present state of the
universe\dots started to evolve from an improbable [i.e. special] state''  \cite{Bol97}. 
This explanation of the arrows of time of our universe governed by time neutral dynamical laws has not changed in over a century,  even as our ideas of state and dynamics have changed. However, throughout, it has mostly been assumed that  the arrows of time point in one direction over the whole of  spacetime {\crd in the absence of special final conditions}. In this paper we show that the fluctuation arrow of time will be bidirectional in the bouncing universes predicted by the no-boundary quantum state \cite{HH83}. 

The visible universe exhibits several time asymmetries called arrows of time. There is the fluctuation arrow defined by the increase in deviations from homogeneity {\crd and the formation of structures such as nucleated bubbles, galaxies, stars, planets, and biota}. There is the arrow defined by the retardation of electromagnetic radiation. There is the psychological arrow defined by our distinction between past, present, and future. There is the thermodynamic arrow {\crf defined by the tendency of presently isolated systems to evolve toward equilibrium in the same direction of time, and the general tendency to increase of an entropy defined by a coarse graining related to conserved quantities.} 

These arrows of time are not independent. As we will discuss briefly below, the electromagnetic, psychological and thermodynamic arrows follow from, or are at least contingent on, the fluctuation arrow\footnote{\crf The authors have written on these other arrows and the relations between them in various places \cite{GH07,Har05}.}.  

The usual story (e.g.\cite{Wei08}) of the fluctuation arrow in an inflationary cosmology starts with an assumed homogeneous, isotropic inflating background spacetime. Fluctuations away from these symmetries are assumed to start at an early time in a state of low excitation like the Bunch-Davies vacuum. The fluctuation modes oscillate until they leave the horizon when their amplitude freezes. Those relevant  for observations today reenter the horizon much later during the  radiation or matter dominated epoch in a state of high excitation relative to the vacuum then. They therefore behave classically and give rise to the fluctuations seen in the temperature of the CMB which grow to become the large scale structures seen today. 

There is an evident arrow of time in this evolution of the perturbations which successfully explains in remarkable quantitative detail many observable features of our universe today. But its very success raises the question of how is it justified in a more fundamental cosmological theory? Why are the classical backgrounds inflating? Indeed, why are there classical histories at all in a theory that incorporates quantum gravity? Why are the fluctuations in a low state of excitation at any time, and, if they are, what time is it? What sets the direction(s) for the growth of fluctuations to operate?  Can there be more than one? We address such questions in this paper in the framework of quantum cosmology.

In the present state of understanding a final theory of our quantum universe consists of two parts. First a theory of dynamics  summarized by a Hamiltonian or action. Second a theory of the universe's quantum state. {\clr There are no predictions of any kind without theories of both.}  The state must play an essential role in any explanation of the arrows of time because the best candidates for theories of dynamics are time neutral  --- $CPT$-invariant in the case of quantum field theory for instance. 

To discuss arrows of time it is necessary to have a well defined notion of time that is not generally available in a quantum theory of spacetime. However, viable theories of the quantum state must predict an ensemble of alternative coarse-grained classical histories at least when the universe is large. Arrows of time are features of this ensemble. In particular we will consider the ensemble of homogeneous and isotropic classical backgrounds and the arrow of time defined by the {\clr quantum} fluctuations about them\footnote{\clt It has been argued (e.g. \cite{Bou11}) that arrows of time can be independent of the quantum state. But these arguments assume eternally inflating classical background spacetimes. In this paper we do not assume a background spacetime but calculate the probabilities for different possible backgrounds from the quantum state as described below.}

The no-boundary wave function (NBWF) \cite{HH83} is perhaps the most developed of the ideas for a theory of the universe's quantum state.  It is attractive for the simplicity of its motivation as the analog of the ground state {\crd for spatially closed cosmologies}. It is naturally connected to the fundamental theory of dynamics.
The protean character of its central topological idea gives hope that it can be extended beyond the semiclassical theory of quantum gravity in which it is currently formulated. Finally, and most importantly,
its predictions are consistent with observations in cosmology {\clt (e.g. as in \cite{HH11})}. The NBWF provides a unified explanation of  the origin of the quasiclassical realm  in a quantum universe \cite{Har09}, the present large size of the universe in Planck units \cite{Haw84}, the approximate homogeneity and isotropy of the observable universe, and the quantum origin of the fluctuations away from these symmetries that are seen in the present large-scale structure \cite{HH85}. It is not free from challenges \cite{Page06} but it is also not completely investigated. This paper concerns its implications for the fluctuation arrow of time. 

The NBWF predictions for the fluctuation arrow were discussed with depth and insight by Hawking, Laflamme, and Lyons (HLL) \cite{HLL93} who also reviewed the earlier history of the subject. They studied the fluctuation arrow in a minisuperspace cosmological model. Geometries were restricted to be spatially closed,  homogeneous, and isotropic with linear fluctuations away from these symmetries. The matter was modeled by a single minimally coupled scalar field and zero cosmological constant. The field was assumed to be homogeneous with linear fluctuations away from homogeneity. 

HLL studied the NBWF predictions for the fluctuation arrow in classical universes that expand from one singularity and then contract to another one. They found that the ensemble of NBWF histories in this class is  time symmetric. However most individual histories are not time symmetric. Rather the fluctuations were small near one singularity and grew throughout the recontraction epoch to be large at the other. The fluctuation arrow of time did not reverse at the moment of maximum expansion but pointed in one direction throughout the cosmological evolution.

This paper extends the HLL results in several directions. First, we will use the same model as HLL but we will consider the complete ensemble of histories predicted by the NBWF. We  base ourselves  on \cite{HHH08a,HHH08b,HHH10a}  where it was found that the NBWF ensemble includes universes with a regular bounce at a minimum radius. These were not treated by HLL but turn out to provide the dominant contribution to probabilities for our observations \cite{HHH10b}. Second, we will include a cosmological constant. Third, we will use a formulation of quantum mechanics \cite{Har95c,HHrules} in which the wave function predicts probabilities for decoherent sets of alternative, coarse-grained, real, Lorentzian histories of the universe. These are distinct from the generally complex {\crj extrema}  of the action used by HLL (although closely related to them). This distinction becomes particularly important in the bouncing universes. We will connect the regularity properties of the complex {\crj extrema}  to the regions of Lorentzian histories where the fluctuations are {\crf in a low state of excitation.} We will find that in bouncing universes, the NBWF predicts that {\cre fluctuations are small at the bounce and that}  the fluctuation arrow of time is bidirectional, extending away from the bounce in both directions\footnote{\crf A bidirectional fluctuation arrow of time was suggested in \cite{HHH08a,HHH08b}. The behavior of the fluctuations was calculated in \cite{HHH10a}. This paper gives the detailed argument.}.

Bidirectional arrows of time for bouncing universes have been discussed by Hoyle and Narlikar \cite{HN64} and Carroll and Chen \cite{CC04} in contexts outside of quantum cosmology. Within quantum cosmology arrows of time have been discussed many times notably by Kiefer and Zeh \cite{KZ95} and in loop quantum cosmology by Bojowald \cite{Boj09} who reaches conclusions for bouncing universes similar to ours.

The outline of this paper is as follows. Section \ref{4dperspective} describes the four-dimensional perspective on arrows of time and the histories that exhibit them. Section \ref{nutshell} briefly reviews the construction of the ensemble of classical histories predicted by the NBWF illustrating that {\clr with} the simple model we consider in this paper. Section \ref{arrows} gives a unified treatment of the NBWF predictions for the fluctuation arrow of time in the various kinds of universes {\crc in the NBWF classical ensemble}.  {\clr Section \ref{through} discusses extrapolating through a bounce.}  Section \ref{otherarrows} describes qualitatively the connection between the fluctuation arrow of time and other arrows of time the universe exhibits. Section \ref{causalinfluences} contains a brief discussion of the implications {\crg for causality of a} bidirectional arrow of time in bouncing universes. Section \ref{before} explains how to answer the question `What came before the big bang?' Some conclusions are in \ref{conclusions}.  An Appendix introduces the machinery necessary to discuss the fluctuations.
 
\section{A Four-Dimensional Perspective on Arrows}
\label{4dperspective}

A common way of thinking about arrows of time is in terms of initial and final states at particular instants of time. The quote from Boltzmann at the start of the Introduction is an instance. For example, if the system's initial state has a low value of a certain coarse grained entropy then that entropy will  increase away from the initial time with high probability. It will continue to increase unless constrained by a final state that makes it decrease (e.g \cite{Coc67,GH93b}). Asymmetries between initial and final conditions are thus one explanation of time-asymmetries of universes governed by time neutral dynamical laws.

This paper takes a more general perspective. We will consider arrows of time as properties of four-dimensional histories of spacetime geometry and matter fields. Fluctuation arrows for instance can be described by giving the spacelike surfaces on which the fluctuations are small and the directions of their increase and decrease away from these surfaces. The surfaces could be initial and final surfaces but need not be. 

There are several reasons why {\crd this}  more general history perspective {\crd on arrows of time}  is useful. At the simplest level a cosmological geometry like a bouncing universe may not have natural spacelike surfaces that define a beginning or end.  Further, four-geometries generally have no preferred foliations by spacelike surfaces that uniquely define a notion of instants of time. 

A deeper reason is that the generalizations of quantum theory that predict probabilities for four-dimensional histories of the universe will not generally have equivalent  3+1 formulations in terms of quantum states evolving unitarily through a preferred foliating family of spacelike surfaces. When spacetime geometry is a quantum variable there is no fixed spacetime geometry to foliate. That is the case for the generalized quantum mechanics that stands behind this work \cite{Har95c,HHrules}. General relativity makes sense as a theory of four-dimensional spacetime even when there is no equivalent 3+1 initial value formulation. Generalized quantum theory makes sense as a quantum theory of {\clr four-dimensional} histories even when there is no equivalent 3+1 formulation in terms of states evolving unitarily through moments of time.

\section{Arrows of time from the No-Boundary State}
\label{nutshell}

The essence of the derivation of the fluctuation arrows of time from the NBWF can be stated very simply. We present it in this section {\crh and the next two}. It needs one technical result to be complete. We present that along with a review of the prerequisite machinery developed in our earlier papers \cite{HHH08a,HHH08b,HHH10a}  in the appendix. 

\subsection{Quantum States and Classical Histories}
\label{sah}
A quantum state of the universe is specified by a wave function $\Psi$ on the superspace of  geometries ($h_{ij}(x)$)  and matter field configurations ($\chi(x)$) on a closed spacelike {\cob three}-surface $\Sigma$. Schematically we write $\Psi=\Psi[h,\chi]$. We assume a cosmological constant $\Lambda$ and a single scalar field $\phi$ moving in a {\clr potential $V(\phi)$ as a model for the matter. The function $\chi(x)$ is the scalar field configuration on $\Sigma$. We assume that  $\Sigma$ has the topology of a three-sphere.}  

{\clr For illustrative calculations  we will use a quadratic potential }
\begin{equation}
V(\phi)=\frac{1}{2}m^2\phi^2 \ .
\label{potential}
\end{equation}  
 {\cre Here and throughout we use Planck units unless explicitly indicated otherwise. The value $m\sim10^{-6}$ leads to fluctuations of the size observed in the CMB. The field value corresponding to a Planck scale energy density $V\sim1$ is thus $\phi\sim 1/m \sim 10^6$.} The observed value of  $\Lambda$ is $\sim 10^{-122}$. 

We assume the no-boundary wave function as a model of the state \cite{HH83}. The NBWF is given by a sum over histories of geometry $g$ and fields $\phi$ on a four-{\clr disk} with one boundary $\Sigma$. The contributing histories match the values $(h,\chi)$ on $\Sigma$ and are otherwise regular. They are weighted by $\exp(-I/\hbar)$ where $I[g,\phi]$ is the Euclidean action along a complex contour chosen so that the defining integral converges and the result is real. (There is more detail in the Appendix.) 

In some regions of superspace the path integral can be approximated by the method of steepest descents. There the NBWF will be approximately given by a sum of terms of the form 
\begin{equation}
\Psi[h,\chi] \approx  \exp\{(-I_R[h,\chi] +i S[h,\chi])/\hbar\} ,
\label{semiclass}
\end{equation}
one term for each complex extremum $(g,\phi)$ of the action $I[g,\phi]$.  Then $I_R[h,\chi]$ and $-S[h,\chi]$ are the real and imaginary parts of the  action, evaluated at the extremum.

In regions of superspace where $S$ varies rapidly compared to $I_R$ (as measured by quantitative classicality conditions \cite{HHH08a}, \eqref{classcond}.) the NBWF predicts that  the geometry and fields behave classically. This is analogous to the prediction of the classical behvior of a particle in a WKB state in non-relativistic quantum mechanics. More specifically the NBWF predicts an ensemble of {\crh spatially closed} classical Lorentzian {\crh cosmological} histories that are the integral curves of $S$ in superspace.  This means that they obey the classical equations relating momenta {\clt  $\pi_{ij}$  and $\pi_\chi$, involving the time derivatives of $h_{ij}$ and $\chi$, to $S(x)$  as follows:}
\begin{equation}
\label{momenta}
\pi_{ij}(x) =\delta S /\delta h_{ij}(x), \quad \pi_\chi(x)\equiv \delta S /\delta\chi(x) .
\end{equation}
The solutions $h_{ij}(x,t)$ and $\chi(x,t)$ to \eqref{momenta}  define field histories by ${\hat\phi}(x,t)\equiv\chi(x,t)$ and Lorentzian four-geometries $\hat g_{\alpha\beta}(x,t)$  by
\begin{equation}
\label{lorhist}
ds^2 = -dt^2 + h_{ij}(x,t)dx^i dx^j \equiv {\hat g}_{\alpha\beta}(x,t) dx^\alpha dx^\beta
\end{equation} 
in a simple choice of gauge\footnote{We follow the notation introduced in \cite{HHH08b} that the complex extrema are denoted by $(g_{\alpha\beta}(x,t),\phi(x,t))$  and the real four-dimensional Lorentzian histories  by $({\hat g}_{\alpha\beta}(x,t),\phih(x,t))$. Occasionally, as above, when we want to emphasize that the Lorentzian histories are integral curves in superspace we will use the notation $(h_{ij}((x,t),\chi(x,t))$ for them.}.
The resulting ensemble of classical histories have probabilities to leading order in $\hbar$ that are proportional to  $\exp[-2 I_R(h,\chi)/\hbar]$, which is constant along the integral curves\footnote{In the terminology used in our other papers these were called bottom-up probabilities to distinguish them from top-down probabilities that are conditioned on our data. Top-down probabilities are relevant for predicting our observations. Bottom-up probabilities are relevant for discussing the probabilities of global features that  the universe may have whether or not there are any observers. The arrows of time are such a global feature.  All probabilities in this paper are bottom-up.}. 

{\clr We call the complex histories of metric and scalar field that extremize the action {\it fuzzy instantons.}
The fuzzy  instantons are not the same thing as the {\clt classical} Lorentzian histories for which they supply probabilities. The metrics of the fuzzy instantons are generally complex --- neither Euclidean or Lorentzian. The metrics of the Lorentzian histories are real and Lorentzian. The geometries of the fuzzy instantons must be regular on the disk $M$ defining the NBWF. The Lorentzian geometries may have singularities like a big bang or big crunch. 

The fuzzy instantons and the {\clt classical}  Lorentzian histories are connected. The fuzzy instantons restrict the possible Lorentzian geometries.  Each Lorentzian history corresponds to a `point' in the classical phase space spanned by $(\pi^{ij}(x),h_{ij}(x))$. But the relation \eqref{momenta} shows that the ensemble of Lorentzian histories lies on a surface in this phase space of half the dimension of the whole.}

{\clr The conditions for classicality may not be satisfied for all degrees of freedom on $\Sigma$. In that case, one can consider some degrees of freedom classically and the others as moving quantum mechanically in the backgrounds supplied by the classical ones. The discussion in this paper will be an example of this.
We will consider classical homogeneous and isotropic background spacetimes in which small fluctuations evolve quantum mechanically. The crucial point again is that to discuss arrows of time at least some degrees of freedom must behave classically to have a well defined notion of time.}

 The NBWF is real by construction \cite{HH83}. This means that for  every complex extremum $(g,\phi)$ contributing to its semiclassical approximation the complex conjugate $(g^*,\phi^*)$ is also a contributing extremum. The latter gives a term in the wave function of the same form as \eqref{semiclass} but with the opposite sign of $S$. Reversing the sign of $S$ means reversing the sign of the momenta in the predicted Lorentzian histories [cf \eqref{momenta}]. For every history arising from one extremum its time reverse will also occur in the ensemble with the same probability since $I_R$ is the same for both. As stressed by HLL, individual classical histories are not generally time symmetric, but the classical ensemble of histories is time symmetric. 
 
Since its ensemble of classical histories is time symmetric, it isn't the NBWF that has an arrow of time. Rather, arrows of time are features of individual histories in the ensemble. As we will argue in Section \ref{otherarrows},  the psychological and other observable arrows are  aligned with the fluctuation arrow. It is only such alignments that are observable distinctions  ---  not the orientation with respect to an arbitrary time coordinate.

\subsection{A Minisuperspace Model}
\label{mss}
We will work in a minisuperspace model in which geometry and fields are restricted to be compact,  homogeneous and isotropic,  with linear fluctuations in geometry and field away from these symmetries. The homogeneous and isotropic three-geometries can be represented in terms of a  scale factor $b$ as
\begin{equation}
\label{scalefactor}
dS^2 = h_{ij}(x) dx^i dx^j = b^2 d\Omega^2_3 
\end{equation}
where $d\Omega^2_3$ is the round metric on the unit three-sphere and $b$ is constant over the sphere. 
A set of coordinates on the minisuperspace consists of the scale factor $b$ of the geometries, the  homogeneous value of the scalar field $\chi$ and gauge invariant measures of the fluctuation modes\footnote{For the details of the fluctuations and their gauge invariant measures see \cite{HLL93} and \cite{HHH10a} especially equation (A1) of that paper that defines the $z$'s.}  $z_{(n)}$, $(n)=(n,\ell,m)$, $n= 2,3,\cdots$ etc.
The wave function is defined on the configuration space spanned by these coordinates, $\Psi=\Psi(b,\chi,z)$. {\clt where we abbreviate by $z$ the whole collection of fluctuations $z\equiv (z_{(1)},z_{(2)},...)$. } {\cred The division into background and fluctuations is thus well defined. }

{\cred Arrows of time are established by the directions in the background time in which the fluctuations systematically grow. Eventually small fluctuations become large making the universe inhomogeneous and the small fluctuation approximation no longer valid. Eternal inflation is one example. Another is  dissipative collapse that produces localized objects like starts and galaxies. We expect, however, that the arrow of time established in the linear regime will continue in the non-linear one in the vast majority of situations.}

To determine the NBWF's predictions for the growth of fluctuations we first find its predictions for the ensemble of homogeneous and isotropic, classical Lorentzian,  background histories $(b(t),\chi(t))$ as described in detail in Section \ref{mss}. Fluctuations away from homogeneity and isotropy in any one of these backgrounds can be described quantum mechanically by a wave function $\psi(b,\chi, z)$ (\eqref{qftwf}). In a particular background  $(b(t),\chi(t))$ this becomes a wave function $\psi(z,t)$ that evolves by the background Schr\"odinger equation with a time-dependent Hamiltonian determined by $(b(t),\chi(t))$.
Thus, a classicality condition (cf \eqref{classcond}) is only enforced for the homogeneous and isotropic backgrounds to give a well defined notion of time with which to discuss the quantum mechanical evolution of fluctuations. The growth of fluctuations is then measured by the correlators of $z_{(n)}$  as function of time in the state of fluctuations specified by their wave function\footnote{\clr Treating fluctuations quantum mechanically or using the term `quantum fluctuations' does not exclude their classical behavior. Classical physics is an approximation to quantum physics. Classical fluctuations are those for which quantum mechanical probabilities are high for suitably coarse-grained correlations in time specified by classical dynamical laws.  Being able to treat the fluctuations quantum mechanically is essential to understanding their behavior in the early universe because a given mode does not become classical until it has left its horizon which can take a number of efolds (e.g. \cite{HHH10a}).}. {\cred The rate of growth of the fluctuations is determined by this state 
[cf \eqref{ptnrsp}]. But it is the {\it direction} of growth that is of interest for the arrows of time. }

In our minisuperspace models the homogeneous and isotropic fuzzy instantons have an $O(4)$ symmetry about the center of the disk which we call the `South Pole' (SP). The metric can be represented
\begin{equation}
\label{nrSP}
ds^2 = d\tau^2 +a^2(\tau) d\Omega_3^2 
\end{equation}
where $a(0)=0$ at the SP $\tau=0$. 

{\crf The simplest and most familiar example has zero scalar field. The instanton is the half of a Euclidean round 4-sphere connected smoothly to Lorentzian de Sitter space at the equator.} In our scalar field model, the NBWF predicts a one-parameter family of homogeneous and isotropic Lorentzian histories, which we label by the absolute value $\phi_0$ of the scalar field at the SP of the corresponding fuzzy instanton. Remarkably all {\clr classical} Lorentzian histories in the ensemble predicted by the NBWF exhibit a period of inflation with a number of efoldings that is approximately given by $3 \p0^2/2$ \cite{HHH08b}.

{\clt The core of the connection between regularity properties of the fuzzy instantons and the regions of Lorentzian histories where the fluctuations  are small is as follows: } 
Any fluctuations away from the $O(4)$ symmetry must vanish at the SP for the perturbed {\crd fuzzy instanton} geometry to be regular there.  Thus, fluctuations must be small in the range of small $\tau$ where the three-geometries also have small volume. It turns out that this property of the complex extrema implies a similar property of the Lorentzian histories that follow from them:  {\it The {\clr classical} Lorentzian histories have one and only one range of three-surfaces where the fluctuations are small.   For bouncing universes this range is at the bounce\footnote{We assume that histories bounce at most once. Classical histories with multiple bounces were exhibited in \cite{Pag84} but  were a set of measure zero.}. For non-bouncing universes this range is near one singularity.} From this property of the fluctuation histories we can deduce the arrows of time as we discuss below. 

To derive this result we show that there is a representation of both the fuzzy instanton and the histories in which the SP of the instanton is  `close' to one range of three-surfaces in each Lorentzian history.  This requires some machinery. We give this in the appendix, and proceed immediately to discuss the implications of this result {\crd for the arrows of time}.  

\section{The Fluctuation Arrows}
\label{arrows}

\subsection{Homogeneous and Isotropic Backgrounds}

The  one-parameter family of  homogeneous and isotropic Lorentzian backgrounds predicted by the NBWF in our model can be divided in two classes by whether the Lorentzian histories {\clr (extended by the classical equations if necessary)} have a bounce or do not. Histories in the bouncing class can have singularities {\clr but need not have any.} Histories in the non-boucing class will have singularities.   A singularity in this paper will mean any region where the energy density in the scalar field exceeds the Planck density.  Semiclassical analysis will not be accurate near a singularity but that is not an obstacle to prediction {\clr in non-singular regions}.  Probabilities for the observable properties of four-dimensional histories at the present time are calculated directly from the fuzzy instantons which are {\clr everywhere regular on $M$} \cite{HHH08a}.

Explicit calculation gives the ranges of the parameter $\phi_0$ corresponding to these classes \cite{HHH08b}. There are no homogeneous and isotropic  histories that satisfy the conditions for classicality 
when the universe is large [cf \eqref{classcond}] for $\p0 \lesssim 1.27$. (This is roughly $10^{-6}$ smaller than the $\p0$ corresponding to Planck energy density {\clr for the value of $m\sim 10^{-6}$ that fits CMB observations.)} The range $1.27 \lesssim \p0 \lesssim 1.55$ is the non-bouncing class. The range  $\p0 \gtrsim 1.55$ is the bouncing class.

\subsection{Fluctuations}

{\crj In these two classes of histories {\clt --- bouncing and non-bouncing --- }, we now consider the consequences for the fluctuation arrows of time of }the property that each Lorentzian history has one and only one range of three-surfaces where the fluctuations are small  --- at the bounce or near one singularity when there is no bounce.

\subsubsection{Fluctuation Evolution}
To connect the fluctuation arrow near the initial singularity with the arrows of time today it is necessary to follow the history of the universe between then and now in detail. 

 Expanding the perturbations in harmonics on the three-sphere one finds that the regularity condition at the SP of the fuzzy instantons implies that the perturbations modes $z_{(n)}$ oscillate in their ground state until their physical wavelength $a/n$ exceeds the horizon size. {\clr Beyond that point} the mode tends to a constant value. 
When a perturbation mode leaves the horizon its {\cre amplitude} freezes \cite{HH85}. The variance {\crf of the probability distribution for the fluctuation amplitudes} depends only on the behavior of the potential for values of $\phih$ near its value at the time the mode leaves the horizon.  The  variance is $\sim \cV^3 (\phih)/\cV_{,\phih}^2$  evaluated at that time where
\begin{equation}
\label{calV}
\cV \equiv \Lambda + (1/2)m^2\phi^2, \quad \cV_{,\phih} \equiv d\cV/d\phih.
\end{equation} 
These and subsequent results are very close to the usual inflationary story assuming a Bunch-Davies vacuum. Here that assumption is derived from  the NBWF.

The {\clr magnitude} of the wave function of a perturbation mode  remains frozen until that scale enters the horizon again in the matter (or radiation) dominated era at a much greater value of the scale factor, even though the equation of state of the universe changes in between. The wave function {\clr of the fluctuations} is in the ground state at the time of horizon exit, but at the time of re-entry it will correspond to a superposition of highly excited, classical states. This is the well-known phenomenon of the amplification of quantum vacuum fluctuations in inflationary universes. It gives rise to a spectrum of classical primordial perturbations with rms  amplitude $Q^2 \sim \cV^3 (\phih)/\cV_{,\phih}^2$, evaluated at horizon exit during inflation. 
In the matter dominated era these classical fluctuations grow under gravitational contraction and eventually evolve into the galaxies and other large-scale structures we observe today.

\subsubsection{The Fluctuation Arrows} 

Fluctuations are small near one singularity in non-bouncing universes.  Explict calculation \cite{HHH10a} shows that the NBWF predicts the onset of inflation near this singularity.  By convention we call this the {\it initial} singularity (the big bang).  The rest of the history may expand forever or recontract and end in another singularity (a big crunch)\footnote{For the range of parameters corresponding to these future possibilities see \cite{HHH10a} especially Figure 12.}.  But whatever is the case, the fluctuation arrow of time points away from this initial singularity throughout the history. It does not reverse on recontraction.
Thus, we recover the results of Hawking, Laflamme and Lyons \cite{HLL93} in an especially clean and transparent way.

In  bouncing histories, the NBWF predicts the fluctuations to be small in the regime of small volume near the bounce. Far from the bounce the history may expand forever or recontract to singularity depending on the value of the cosmological constant\footnotemark[7]. 

\subsubsection{Through a Bounce}

{\cls This discussion of fluctuation arrows concerns the classical histories in a region of superspace where the semiclassical approximation \eqref{semiclass} holds and the classicality conditions \eqref{classcond} are satisfied. This region is bounded by small values of $b$ either  because there is a singularity or there is a bounce where $S$ varies slowly. But the past evolution of the universe does not stop there. It might continue to another region of larger $b$ where the classicality conditions are satisfied --- either classically and deterministically or quantum mechanically and probabilistically. If the ensemble of classical histories in that region is the same as on the first side then fluctuations would increase away from the bounce on both sides. The inflationary expansion away from the bounce would give rise to a spectrum of classical primordial perturbations leading ultimately to the formation of a large-scale structure of galaxies, stars, planets, etc  and more complex structures such as biota, civilizations, on both sides. There would be a bidirectional arrow of time.

We will not discuss continuation through singular regimes \cite{SING}. The best chance for a manageable passage is with universes that bounce well above the Planck scale. We discuss these in the next section. 

}

\section{The Bidirectional Arrows of Time of Bouncing Universes}
\label{through}

\subsection{Classical Bouncing Backgrounds}

\begin{figure}[t]
\includegraphics[height=2in]{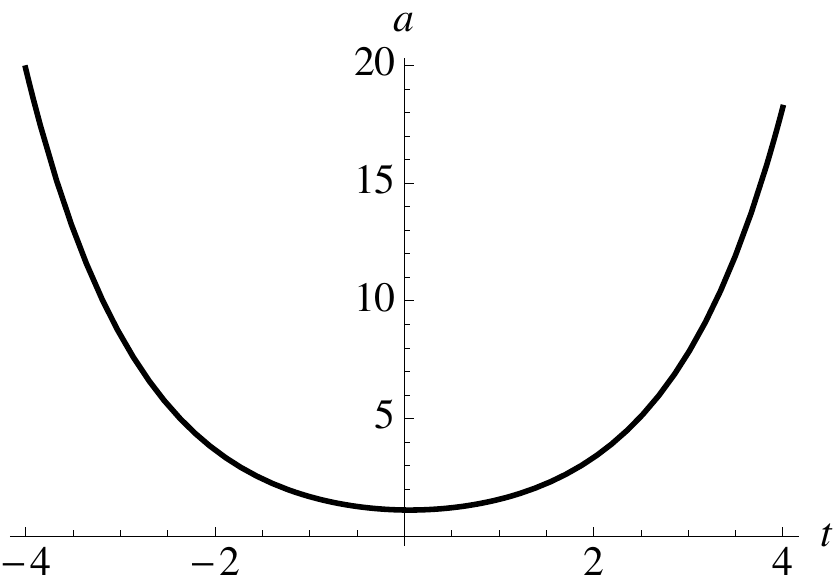}\hfill
\includegraphics[height=2in]{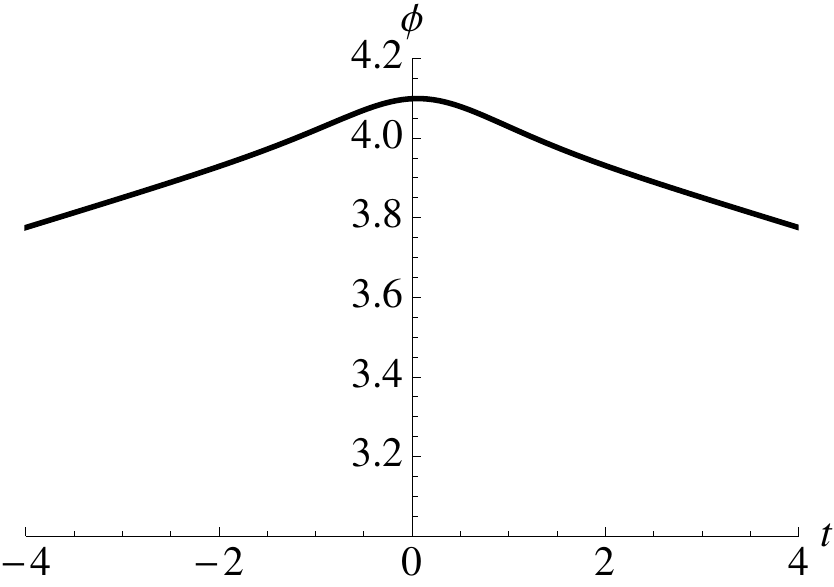}\\
\caption{A bouncing universe.  This figure shows the scale factor $\ah$ (top) and the scalar field $\phih$ (bottom) for a homogeneous, isotropic Lorentzian history in the NBWF classical ensemble. The example shown has $\p0=4$ for $m=.05$.  The quantities $\ah$, $\phih$ and $t$ are all in Planck units. The solution was found by extrapolating the asymptotically real behavior of the $\p0=4$ fuzzy instanton, where the semiclassical approximation applies, backward in time using the classical equations of motion. {\crh The universe inflates on both sides of the bounce.}}
\label{bouncingmodels}
\end{figure}

Suppose that for bounces that occur at a volume sufficiently above the Planck scale the backgrounds $(b(t),\chi(t))$ can be continued through a bounce by solving the classical equations of motion. That is physically plausible and supported in simple non-relativistic models. (See the discussion in Section \ref{bcu}.) Figure \ref{bouncingmodels} shows some representative examples in the minisuperspace models under discussion. 
The backgrounds are not generally time symmetric, but the time  asymmetry is small for the large $\phi_0$ histories that dominate the ensemble \cite{HHH08b}.

\begin{figure}[t]
\includegraphics[height=2.2in]{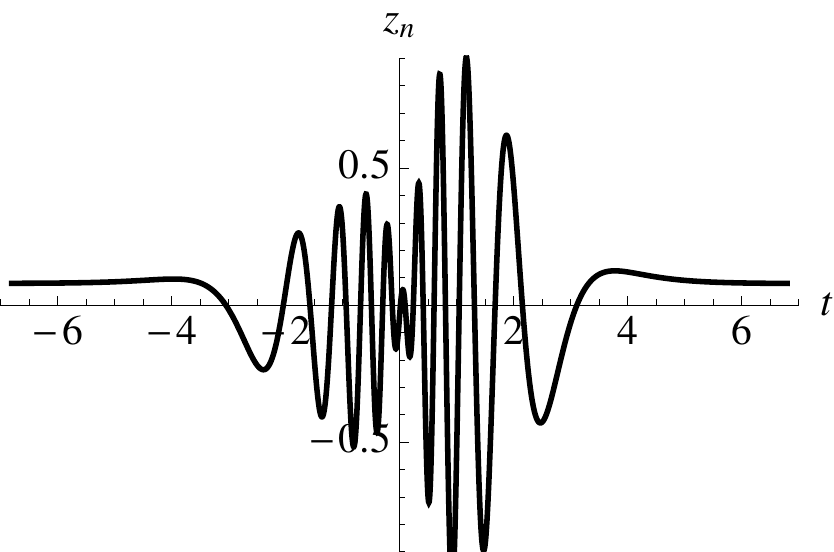} 
\caption{{\crc Fluctuations in a bouncing universe.} This figure shows the behavior of the gauge invariant fluctuation mode $z_{(n)}$ with wavenumber $n=20$, as a function of time {\clr in Planck units} in the bouncing background of Figure \ref{bouncingmodels}. The fluctuation history was calculated by {\crf starting with} the behavior of the $n=20$ perturbation mode around the fuzzy instanton at large $t$  {\crf and extrapolating that}  backwards in time using the Lorentzian perturbation equations. This solution determines the wave function of the perturbation {\cre to leading order in $\hbar$.} The fluctuation oscillates in its ground state in the regime of three-surfaces near the bounce at $t=0$ when the size of the universe is small. The fluctuation  amplitude freezes  when the physical wavelength of the mode becomes larger than the horizon size. This happens for the mode shown around $t= \pm 4$ on either side of the bounce.  The fluctuation modes will enter the horizon again  much further away from the bounce, on both sides, as classical perturbations with an expected amplitude $Q^2 \sim \cV^3 (\phih)/\cV_{,\phih}^2$ evaluated at horizon exit during inflation. During matter domination perturbations continue to grow in the two directions away from the bounce {\clr (on time scales much larger than illustrated here)}, leading to a bidirectional fluctuation arrow of time in the bouncing histories predicted by the NBWF.}
\label{fluctuations}
\end{figure}

\subsection{Fluctuations in a Bouncing Background}

{\clr Focus attention on a particular background $(b(t),\chi(t))$ that has been evolved classically through a bounce. The Schr\"odinger equation in that  background evolves the wave function for the fluctuations 
$\psi(z, b(t), \chi(t))$ through the bounce. The evolved wave function on the far side gives the NBWF predictions for  the amplitude of fluctuations there. The fluctuations under discussion here are small and have quadratic actions [cf. \eqref{pertaction}]. For them, the Schr\"odinger propagator between two different values of $z_{(n)}$ at two different times is specified to leading  order in $\hbar$  by the classical Lorentzian evolution between these values.  This solution is shown in Figure \ref{fluctuations} for the background in Figure \ref{bouncingmodels}.  The figure shows that the fluctuations behave similarly on both sides of the bounce. They oscillate at the bounce, eventually freeze, and  grow away from the bounce  when they enter the horizon at much later times thus defining a bidirectional fluctuation arrow of time. 

}

\subsection{Eternal Inflation}

 We have seen that in bouncing universes, perturbations are small at the bounce and increase away from it on either side. The resulting classical perturbations have a Gaussian distribution with variance $\cV^3 (\phih)/\cV_{,\phih}^2$ evaluated at horizon crossing. This means that in histories with a regime where $\cV^3(\phih)>\cV^2_{,\phih}$, significant probabilities are predicted for {\it large} fluctuations that leave the horizon while this condition holds. This is called the regime of eternal inflation, which in our model occurs in histories where  $\phih (t) \gtrsim1/\sqrt{m}$ near the bounce. 

The NBWF thus predicts that histories of this kind develop large inhomogeneities {\it on both sides of the bounce} on the very large scales that left the horizon in the regime of eternal inflation \cite{HHH10a}. These scales correspond to superhorizon scales at the present time and are thus not directly observable.

 Large inhomogeneities make it difficult to estimate our distance in time from the regime of small fluctuations near the bounce. However, this does not alter our finding that the fluctuation arrow of time  reverses at the bounce whether or not the history exhibits regimes of eternal inflation. 

The arrow of time of the perturbations on very large scales predicted by the NBWF has further interesting implications in more complicated models where the potential has one or several positive false vacua or sufficiently flat extrema. Near each extremum  the NBWF predicts a set of (eternally) inflating universes that exhibit a bounce. Quantum fluctuations around these histories occasionally lead to the nucleation of `pocket' universes that are like bubbles of true vacuum  which expand in an eternally inflating region of false vacuum. It is usually assumed that all the pocket universes form in the same direction of time. In our framework this assumption is justified as a consequence of the fluctuation arrow of time predicted by the NBWF. Since the fluctuations are small near the bounce we predict there are no bubbles early on. {\clr No bubbles can nucleate until  universe is bigger than the horizon size set by the false vacuum energy.} Further, since the fluctuation arrow is homogeneous in space we expect that in the absence of a special final boundary condition, all pocket universes on a given side of the bounce emerge in the same direction of time but in opposite directions on opposite sides.

\section{The other arrows of time}
\label{otherarrows}

The fluctuation arrow(s) of time can be seen as the origin of some other arrows  that the universe exhibits. We will describe very briefly {\crc and qualitatively} the fluctuation arrow's connection with the electromagnetic arrow, the thermodynamic arrow, and the psychological arrow of time\footnote{For more details see, e.g. \cite{Har05}.}.

{\it The electromagnetic arrow of time:} Maxwell's equations are time neutral but free electromagnetic radiation tends to increase in one direction of time. That is the electromagnetic arrow. Free electromagnetic radiation is an example of a fluctuation of a homogeneous and isotropic universe. {(The only homogeneous} electromagnetic field is zero.) Free electromagnetic radiation will be small when the fluctuations are small and the universe is small. {\cre Therefore, most of the radiation we detect is produced from accelerated charges as the universe expands. The electromagnetic arrow thus coincides with the fluctuation arrow; { indeed, it is a part of it}. If we call the direction towards the regime of small fluctuations `the past', observed electromagnetic radiation is retarded; its sources are in the past.

{\it The thermodynamic arrow of time:} {\cre One aspect of the} second law of thermodynamics is the tendency to increase of the total entropy of matter defined by a coarse-graining related to the quasiclassical variables that define the quasiclassical realm \cite{GH07}. To this entropy can be added the entropy produced by the formation of black holes giving the generalized second law of thermodynamics. The thermodynamic  arrow of time is in the direction of the increase of these entropies\footnote{\cre The other aspect of the second law is the homogeneity of the thermodynamic arrow. The individual entropies of presently isolated systems increase mostly in the same direction of time. That is true trivially in the present set of models where homogeneous backgrounds are assumed.}.

As Boltzmann said in the quote at the start of the paper, the second law of thermodynamics can be explained if the universe was in a state of low entropy at one time. But matter in a nearly homogeneous state is in a state of relatively low entropy because gravitational clumping (the growth of fluctuations) increases entropy. Small fluctuations grow, collapse, heat, dissipate energy in retarded electromagnetic radiation, and occasionally produce black holes. In short the growth of fluctuations leads to the increase in entropy\footnote{For a discussion of how much the entropy increases see e.g. \cite{LB90}.}.  The thermodynamic arrow of time is thus in the same direction as the fluctuation arrow.

{\it The psychological arrow of time:} The psychological arrow of time is the sharp distinction between past, present, and future made by human information gathering and utilizing systems. We remember the past, experience the present, and predict the yet to be experienced future.  This psychological arrow can be understood as arising from the electromagnetic arrow and the thermodynamic arrow in a model of how humans process temporal information e.g. \cite{Har05}. Merely the fact that we receive electromagnetic signals from the past and not the future is {\cre a significant} part of the argument. The psychological arrow is thus in the same direction as the fluctuation arrow. 

\section{Causal Influences}
\label{causalinfluences}

{\crj In the present, information about the past is more accessible than information about the future. To know  how galaxies were distributed long ago we have only to observe the light from them today.  But to know something about how galaxies will be distributed in the far future we have to do a calculation. Put differently, in the present there are more accessible records correlated with events in the past than there are of events in the future. This asymmetry in correlation is often expressed very loosely by saying that we are influenced by events in the past but not the future --- a causal arrow of time.

The retardation of electromagnetic radiation is one reason for this asymmetry in correlation. But, as is well known, the asymmetry is also consistent with the second law of thermodynamics. If we had as much information about the future as the present the missing information measured by an appropriate entropy would not increase. In short, the asymmetry in correlation arises from the arrows of time that are connected to the fluctuation arrow. 

In bouncing universes  the causal arrow will point with the fluctuation arrow in opposite directions on opposite sides of the bounce. This means that events now will have little effect on the far side of the bounce because it is in our past. Conversely events there can have little influence on us {\clt because we are in their past}. {\clr The similar development of fluctuations on the far side mean that it should contain its own galaxies, stars, planets and biota.} But we have no more chance of receiving a message from observers on the far side of the bounce than we have of sending a message back in time with instructions on how to avoid a turn of events  that had unfortunate consequences later.  Cosmological events such as symmetry breaking or black hole formation on the far side of the bounce can therefore be expected to have little effect on our observations. 
}

{\clr It may be of intellectual interest to calculate probabilities for various events on the other side of a bounce from the one where we are. But the results of those calculations will be of little help in predicting our observations. The situation is similar that with events in the far future. Elementary causality implies that our observations are unaffected by future events, rather they depend on events in our past. Further, as the present discussion shows, that region of influence only extends to the bounce if we live in universe that has one. In limiting regimes of causal influence a bounce is not that much different from a singularity. To calculate observations for our observations we can coarse grain both over our future (e.g \cite{HHH10b}), and over the far side of a bounce in our past.}

This is in sharp contrast with the causality in ekpyrotic models of cosmology \cite{ekpyrotic}, pre-big bang models  \cite{prebb},  and in Penrose's {\crj conformal} cyclic cosmology \cite{Penrose}. {\crh In these}  a smooth, ordered state in the infinite past {\crh is assumed. As the universe}  evolves, departures from this state grow in time. In pre-big bang models, {\crh for instance},  the large-scale structures we observe today originate from small perturbations in a contracting phase that precedes the current phase of expansion. { Typically the transition between contraction and expansion is} classically singular. Pre-big bang models rely on the assumption that evolution continues through singularities and, furthermore, that the quantum mechanical evolution across the bounce is essentially classical, relating the  quasiclassical realms on both sides. Perturbation modes relevant for observations today are argued to remain frozen across this transition which means they carry information across the bounce. The fluctuation arrow of time therefore points in the same direction across the entire spacetime in cosmologies of this kind.

\section{What Came Before the Big Bang?}
\label{before}
Anyone who has every delivered a public exposition of contemporary cosmology will be familiar with this question. The answer given by NBWF quantum cosmology has been {\crf the subject of}  this paper.  

Observations of the CMB and consistency of the theory of early universe nucleosynthesis {\crf provide} ample evidence that {\crh early on}  the temperature of our universe must have been high enough to disassociate nuclei. We call that epoch the `big bang'. As illustrated by the models in this paper, the universe could have evolved through this epoch essentially classically never producing a singularity. In that case there was a history of contraction to this bounce before the big bang. However, {\crf as discussed above,} events in that  epoch have no causal effect on our observations because the fluctuation, electromagnetic, and thermodynamic arrows of time point in the opposite direction of ours.  We can ignore the period before the big bang for predictive purposes.

On the other hand present data is consistent with a classical singularity in the past. {\clr Near the singularity}  there is no longer a high probability of correlations in time characterizing classical spacetime. {\clr Without}  a classical spacetime to define `before' and `after' the question of what came before the big bang is no longer meaningful, At singularities the classical notion of time breaks down. Whether history can be extended quantum mechanically through classical singularities is still an open question \cite{SING}. 

\section{Conclusion}
\label{conclusions}

Arrows of time in universes governed by time-neutral dynamical laws can often be understood as asymmetries between initial and final conditions. The standard story of the growth of structure in inflationary cosmology {\crf sketched in the introduction is an example. 
Similarly the growth of structure in pre-big-bang models \cite{prebb} also arises from asymmetries between initial and final conditions. 

Notions of `initial' and `final' presuppose a background classical spacetime in which to locate them. But in a quantum theory of gravity fixed spacetime geometries cannot be presupposed. Rather, they must be derived as predictions of the quantum state for suitably coarse-grained alternatives for the universe's four-dimensional history. Then it is only natural to derive arrows of time as aspects of the four-dimensional histories predicted by the state.  Arrows of time then emerge directly from a fundamental quantum description of the universe along  with classical spacetime. That is the approach that we have followed in this paper in simple models incorporating the no-boundary quantum state. 

We showed for the bouncing universes predicted by the NBWF that fluctuations are in a low state of excitation near the bounce and that the fluctuation arrow of time points in opposite directions on opposite sides of the bounce. This would require extraordinary fine-tuning in a theory where initial and final conditions on the fluctuations are imposed at the large ends of the universe. But the result emerges very naturally in a quantum theory of four-dimensional histories. {\clr Any discussion of fine-tuning requires a measure. In the measure supplied by the NBWF a bidirectional arrow of time in bouncing universes is not fine-tuned.}

 The arrows of time of our classical universe are central to our experience and even to our existence. It is striking to think that  these everyday asymmetries of the world emerged 14 Gyr ago, and have remained pointing in the same direction since, as a consequence of the universe's quantum state.

}
\vskip .2in

\noindent{\bf Acknowledgments:}  { We thank Stephen Hawking for many conversations and early collaboration on this problem.} We thank Chris Pope, Sheridan Lorenz and the Mitchell Institute for several meetings at Cooks Branch.  {\clr We thank M. Kleban, D. Page, M. Poratti, and A. Vilenkin for critical discussions.} The work of JH was supported in part by the US NSF grant PHY07-57035 and  by Joe Alibrandi. The work of TH was supported in part by the Agence Nationale de la Recherche (France) under grant ANR-09-BLAN-0157.

\appendix
\section{What Fuzzy Instantons Imply about Lorentzian Fluctuations}
\label{app}

This appendix is devoted to showing that the NBWF implies the behavior of the fluctuations in bouncing and non-bouncing universes assumed at the end of Section \ref{sah}. We shall assume the prescription summarized there for the calculation of the classical ensemble and its probabilities. For background consult  Sections II and III of  \cite{HHH08a} and Section III of \cite{HHH10a}. Henceforth, we call these papers CU and EI respectively. 

\subsection{The Minisuperspace Model and the NBWF}
\label{mss}
Our models have geometry coupled to a single scalar field $\phi$ moving in a quadratic potential $V=(1/2)m^2\phi^2$ and a cosmological constant. 
We consider minisuperspace models defined by linear perturbations away from closed, homogeneous and isotropic three-geometries and field configurations. Minisuperspace is spanned by the scale factor $b$ of the homogeneous three-geometries, the homogeneous value of the scalar field $\chi$  and the parameters defining the perturbation modes. We denote the latter collectively by $z=(z_{(1)},z_{(2)},...)$; they are defined precisely in \cite{HHH10a}.  Where convenient, we denote the whole set of coordinates by $q^A$. Thus, $\Psi=\Psi(b,\chi, z) \equiv \Psi(q^A)$ . 

The NBWF is defined by an integral of the exponential of minus the Euclidean action $I$ over complex  four-geometries and field configurations that are regular on a four-disk $M$ with a three-sphere boundary $\Sigma$ on which the four-dimensional histories take the real values $(b,\chi,z)$ \cite{HH83,Haw84}. Schematically we can write
\begin{equation}  
\Psi(b,\chi,z) =  \int_{\cal C} \delta a \delta \phi \delta\zeta  \exp(-I[a(\tau),\phi(\tau),\zeta(\tau)]/\hbar) .
\label{nbwf}
\end{equation}
Here, $a(\tau)$ and $\phi(\tau)$ are (complex) histories of scale factor and scalar field defining a homogeneous, isotropic background. The quantities $\zeta(\tau)=(\zeta_{(1)}(\tau),\zeta_{(2)}(\tau), \cdots)$ denote histories of fluctuations away from homogeneity and isotropy in both metric and matter field. $I[a(\tau),\phi(\tau),\zeta(\tau)]$ is the Euclidean action. The integral is over geometries and matter fields  that  are regular  {\clr on $M$ and such that on the boundary $\Sigma$ the functions $a(\tau)$, $\phi(\tau)$, and $\zeta(\tau)$ take the values $b$, $\chi$, and $z$ respectively. }The integration is carried out along a suitable complex contour ${\cal C}$ which ensures the convergence of \eqref{nbwf} and the reality  of the result \cite{HH90}. 

We restrict to linear fluctuations; only up to quadratic terms  in $\zeta$ are retained in the action in \eqref{nbwf}:
\begin{equation}
I = I^{(0)}[a(\tau),\phi(\tau)] + I^\epb [a(\tau),\phi(\tau),\zeta(\tau)]  .
\label{pertaction}
\end{equation}
Then  $I^{(0)}$ is the action for  the homogeneous isotropic background  and $I^\epb$ is the  action for the linear  perturbations away from that background.

The ensemble of classical histories for this model was calculated in CU and EI. Only a small amount of that discussion is necessary to understand the NBWF predictions for the fluctuation arrows of time. We will summarize that very briefly here but for greater depth and detail the reader is referred to CU and EI.

\subsection{Fuzzy Instantons}
\label{fuzzy}

In some regions of superspace one or more of the integrals in \eqref{nbwf} may be well approximated by the method of steepest descents. We suppose first that this is the case for the integrals over the homogeneous parts of geometry and field and that the back reaction of the fluctuations is small. The wave function will then be approximately given by 
\begin{equation}
\Psi(b,\chi,z) \approx \exp\{[-I^{\epa}_R(b,\chi) + i S^{\epa}(b,\chi)]/\hbar\} \psi(b,\chi,z),
\label{semiclassback}
\end{equation}
one such term for each {\crf complex} history $(a(\tau),\phi(\tau))$ that extremizes the action $I^\epa$, matches $(b,\chi)$ at the boundary of the disk, and is regular elsewhere. For each contribution $I^{(0)}_R (b,\chi)$ is the real part of the action $I^\epa[a(\tau),\phi(\tau)]$ evaluated at the extremizing history and $-S^\epa(b,\chi)$ is the imaginary part. The wave function $\psi$ {\clr of the fluctuations in the homogeneous isotropic background $(a(\tau),\phi(\tau))$} is defined by the remaining integral over $\zeta$
\begin{equation}  
\psi(b,\chi,z) \equiv  \int_{\cal C}\delta\zeta  \exp(-I^\epb[a(\tau),\phi(\tau),\zeta(\tau)]/\hbar) .
\label{qftwf}
\end{equation}

The geometry of the homogeneous isotropic extrema can be written in a suitable set of real coordinates as 
\begin{equation}
ds^2=N^2(\lambda) d\lambda^2 +a^2(\lambda) d\Omega^2_3,
\label{eucmetric}
\end{equation} 
where $ d\Omega^2_3$ is the round metric on the unit three-sphere [cf. CU(3.1)].  The {\crf homogeneous} scalar field is $\phi(\lambda)$. The equations determining the extrema are CU(4.5).  The solutions must be regular on the disk and match the $(b,\chi)$ on its boundary. We can think of $\lambda$ as a radial coordinate on the  disk which ranges from the South Pole (SP) of the geometry at $\lambda=0$ where $a(0)=0$  and $\phi(0)\equiv \p0\exp(i\theta)$  to the boundary at $\lambda=1$ where $a(1)=b$ and $\phi(1)=\chi$.  These boundary conditions determine a one parameter family of  complex extrema. The absolute value of the field at the SP $\p0=|\phi(0)|$ is a convenient choice for the parameter.

The action evaluated on one of these solutions is independent of the metric coefficient $N(\lambda)$  which can therefore be chosen to be an arbitrary complex function. This freedom can be expressed differently by rewriting the equations for the  extrema in terms of a  parameter $d\tau \equiv N d\lambda$. The freedom in the choice of $N$ then corresponds to a freedom in the choice of curve in the complex $\tau$-plane on which the equations are solved.  The possible contours for solving the equations and evaluating the action run from $\tau=0$ at the SP to a complex value $\upsilon$ marking the boundary of the disk. More generally, an extremum can be thought of as a pair of complex analytic  functions $a(\tau)$ and $\phi(\tau)$. Their values along any contour in the $\tau$ plane are a representation of the solution. We shall exploit this in what follows. 

A useful contour  to represent complex extrema  in $\tau=x+iy$ turns out to run along the real-$\tau$ axis from the SP at $\tau=0$ to a value $X$ and then in the imaginary direction $y$ (cf. Figure \ref{contour}).  We call this broken contour $C_B(X)$. For each  $\p0=|\phi(0)|$, one can find an  $X$  and a phase of $\phi(0)$ such that the $a(\tau)$ and $\phi(\tau)$ approach real values at large $y$ along the vertical part of the contour\footnote{As was shown approximately analytically by Lyons \cite{Lyo92}.}. In this case the real part of the action $I_R(b,\chi)$ becomes nearly constant at sufficiently large $y$. In particular $I_R$ then varies slowly compared to $S$ and the conditions for classicality [cf. CU(3.13), \ref{classcond}] are satisfied. The real asymptotic behaviors of $a(\tau)$ and $\phi(\tau)$ can be interpreted as part of a real Lorentzian history $(\ah(t),\phih(t))\equiv (b(t),\chi(t))$ in which imaginary $\tau$ corresponds to real Lorentzian time $t$. To leading order in $\hbar$, he probability of this history is the asymptotic value of  $\exp(-2I_R/\hbar)$.  This choice of contour therefore gives a representation both of the saddle point and of the Lorentzian history it predicts. In particular it makes manifest the connection between them\footnote{In the simple example of no scalar field the part of the solution along the real axis corresponds to the surface of a real Euclidean 4-sphere. This is joined across the equator at $X=\pi/2$ to a real Lorentzian de Sitter space along the vertical part of the contour. In this {\crf special} case the extremum is real. But  in general {\cre the extrema} are complex --- fuzzy instantons --- as we have described above.}. We will use this in what follows.  

 As mentioned in \ref{sah},  the equations for the complex extrema are real analytic meaning that if they are solved by $[a(x+iy),\phi(x+iy)]$ then the complex conjugate of these functions is also a solution that can be written as $[a(x-iy),\phi(x-iy)]$. This corresponds to the contour in Figure \ref{contour} reflected about the real axis. 
 
 {\clr When the wave function of fluctuations defined in \eqref{qftwf} is evaluated on a particular homogeneous and isotropic Lorentzian background $(b(t),\chi(t))$, it gives a wave function $\psi(z,t)$ describing the evolution of the quantum fluctuations in that background. This wave function satisfies the Schr\"odinger equation with a Hamiltonian determined\footnote{See e.g. EI, Section III and the references to earlier literature therein.} by $(b(t),\chi(t))$. That Hamiltonian is quadratic in the $z$'s and their derivatives. The Schr\"odinger propagator between two different $z$'s at two different times is proportional to $\exp(iS_{\rm cl})$ ,where $S_{\rm cl}$ is the action of the Lorentzian history connecting those points. This Lorentzian history provides a convenient summary of the dynamics. An example is illustrated in Fig \ref{fluctuations}.
 
 }

\subsection{Bouncing Classical Universes} 
\label{bcu}

The semiclassical approximation to the NBWF will hold in those regions of superspace where the action $I(q^A)$ varies rapidly. Classical histories can be reliably inferred only in those parts of that region where the stronger classicality conditions hold. These say that the imaginary part of the Euclidean action $S(q^A)$ varies rapidly compared to the real part $I_R(q^A)$. {\crf More precisely  the classicality conditions are [cf. CU (3.13), (3.17)]}
\begin{equation}
\label{classcond}
|\nabla_A I_R| \ll |\nabla_A S|,  \quad  |(\nabla I_R)^2| \ll |(\nabla S)^2| 
\end{equation}
{\crf where the norms are constructed with the superspace metric.} The integral curves of $S$ then define Lorentzian histories\footnote{\clr The coarse graining generally required for classicality has not been discussed here.  But once we have a classical background we treat the fluctuations at a fine-grained level quantum mechanically.}.

The classicality conditions do not hold in regions of superspace where the classical universes bounce { because  $S$ varies slowly there. We found in
\cite{HHH08b} that for the most interesting case of  large $\phi_0$ this region is small.}   Intuitively we do not expect that classical predictability fails in a universe that bounces at a radius well above the Planck scale any more than we expect it to fail in a region of maximum expansion in a recollapsing universe where $S$ also varies slowly. That intuition is supported in simple non-relativistic models.

\begin{figure}[t!]
\includegraphics[width=1.3in]{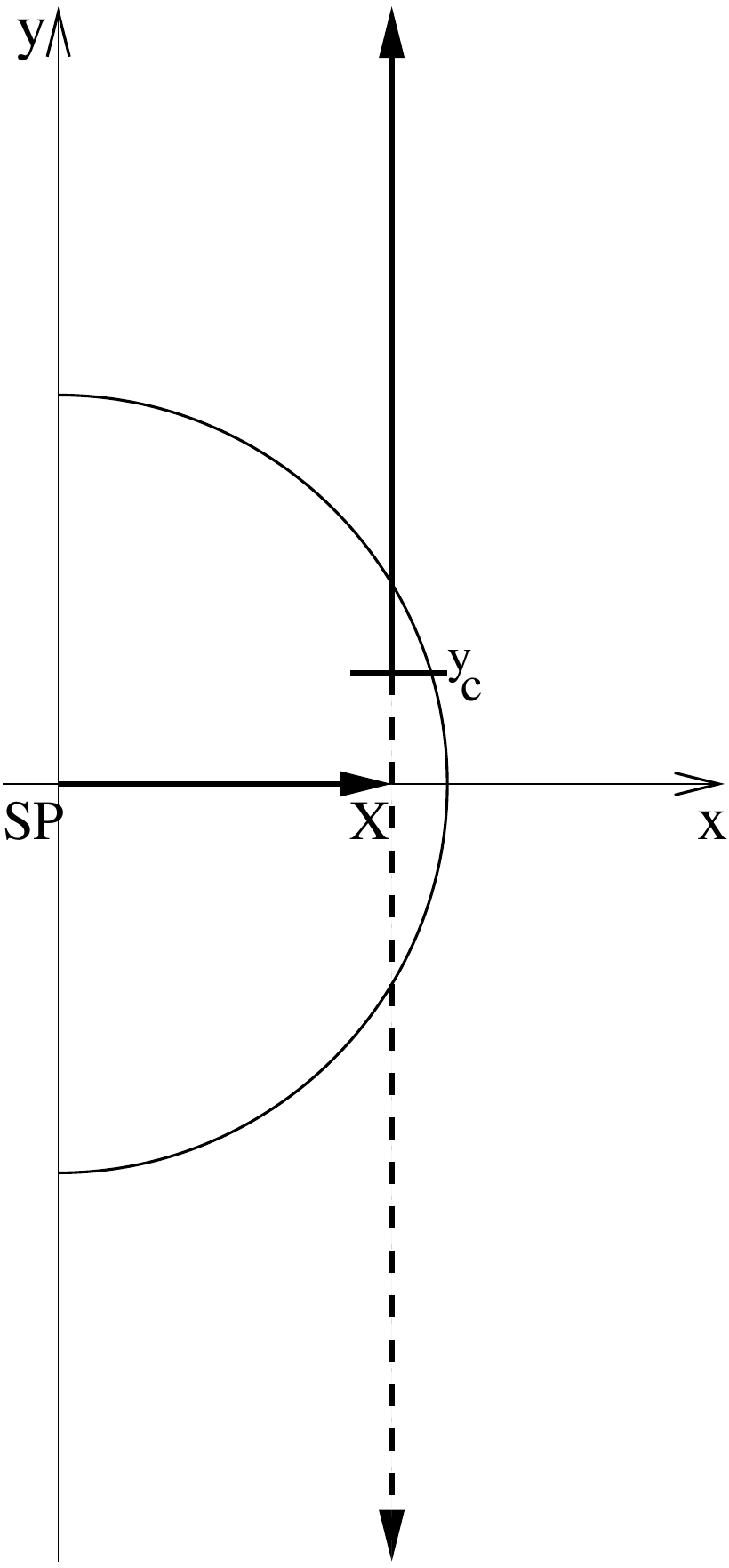}  
\caption{A contour in the complex $\tau$-plane for representing both a complex saddle point and the Lorentzian history it predicts. The figure shows the contour $C_B(X)$ that starts along the real axis from the SP at $\tau=0$, breaks at $X$, and extends upward in the imaginary $\tau$ direction. The complex saddle points start at the SP  with $a(0)=0$, $\phi(0)\equiv \p0 \exp(i\theta)$ and conditions of regularity. By tuning $\theta$ and $X$ for each $\p0$ a vertical contour can be found along which the imaginary parts of $a(X+iy),\phi(X+iy)$ { become negligible beyond some value $y_c$, which decreases for increasing $\p0$. { The variation in $I_R$ also becomes negligible so that the classicality conditions are satisfied.} For large $\phi_0$ one has $y_c \ll 1$. The approximately real values of $a$ and $\phi$ for $y > y_c$ provide Cauchy data for the classical Lorentzian `background' history $(\ah(t),\phih(t))$ predicted by the saddle point, which can therefore be labeled by $\p0$. The Lorentzian histories can be extrapolated classically to values lower than $y_c$ as indicated by the dotted line.} For low values of $\p0 \sim {\cal O}(1)$ the histories are singular in the past. However for larger values of $\p0$ the histories exhibit a bounce in the past at a finite radius $b$ at $y \approx 0$. In the latter case we assume one can reliably extrapolate along the vertical part of the contour to negative values of $y$ where we find the history enters another inflationary regime. Regularity at the SP of the fuzzy instantons requires that the saddle point fluctuations vanish there. Saddle point fluctuations will therefore be small in a region around the SP indicated by the circle. {\crc We find this region includes the part of the contour up to $y_c$. The predicted Lorentzian histories of fluctuations coincide with the saddle point fluctuations along the solid part of the contour. The wave function of the fluctuations can be extrapolated along the $x=X$ line using the perturbation equations [cf. EI(A2)]. We find the Lorentzian fluctuations are small in the regime near $t \sim y_c$  where the universe is also small.  For bouncing universes fluctuations oscillate in their ground state near the bounce {\crf as illustrated in Figure \ref{fluctuations}  and eventually } increase in both directions away from there indicated by the arrows. This gives rise to large-scale structures  on both sides of the bounce. Hence the bouncing universes predicted by the NBWF exhibit a bidirectional fluctuation arrow of time.}} 
\label{contour}
\end{figure}
The semiclassical form \eqref{semiclassback} and the classicality conditions are {\it sufficient} criteria for classicality. In principle the probability for any history can be calculated from the wave function $\Psi(q^A)$ without a semiclassical approximation. {\it We will assume that once classical histories have been identified in a region of minisuperspace where the classicality condition holds they can be extrapolated  to regions where it does not hold using the  classical equations of motion until they become classically singular.} That is an assumption which can in principle be checked in the full quantum mechanical theory\footnote{\clr When the classicality conditions \eqref{classcond} are satisfied the classical histories can be computed either using the classical equations of motion or as the integral curves of $S$. However when \eqref{classcond} are not satisfied, as near a bounce, these notions are different. We thank D. Page for a discussion on this point.}.

We can think of this extrapolation as taking place along a vertical contour in the complex $\tau$-plane. This is illustrated for a bouncing universe in Figure \ref{contour}. There the complex saddle point approaches a real Lorentzian history where the classicality conditions are satisfied in the solid part of the vertical contour but can be extended by classical extrapolation to {\crf the dotted part of the line. In this way we construct classical homogeneous, isotropic backgrounds that contract from a large radius, bounce, and expand again to larger radii   --- bouncing universes.}

\subsection{Fluctuation Saddle Points}
\label{fsp}

At a given homogeneous and isotropic saddle point the remaining integral over fluctuations in the sum \eqref{qftwf} can also be done by the method of steepest descents. In fact, since the integral is Gaussian the steepest descents approximation is essentially exact. 

The complex histories of fluctuations $\zeta(\tau)$ defining the relevant saddle points can be thought of as perturbations of the  background, homogeneous and  isotropic fuzzy instanton. The $\zeta(\tau)$ are solutions to the differential equations determining the extrema. They are determined by the boundary conditions that they match the $z$'s on the boundary of the disk and are regular everywhere inside.  The key point for our arguments is that {\it regularity implies that fluctuations in geometry and field vanish at the SP}, as we now show. 

To exhibit what we need explicitly we use the formalism for the metric fluctuations developed in HLL \cite{HLL93} and employed in EI. We exhibit explicit expressions only for perturbations that are scalars under the $O(4)$ invariance of the  background geometry, but the tensor perturbations work out in a similar manner. It is convenient to work in the (so called) Newtonian gauge where the scalar metric perturbations have the form [cf. EI(4.3)]
\be
ds^2 = d\tau^2 + a^2(\tau) [1-2\psi(\tau,x^i)] d\Omega^2_3 
\label{pmet}
\ee
and the $x^i$ are coordinates on the unit round three-sphere. 
The fluctuations in the scalar field $\delta\phi(\tau,x^i)$ and in the metric $\psi(\tau,x^i)$ can then be expanded in $O(4)$ harmonics $Q^n_{\ell m}(x^i)$ on the three-sphere [cf. EI(4.4),(4.5)]: 
\begin{subequations}
\label{expnharmonics}
\begin{align}
\label{harmonics}
\psi(\tau,x^i) &=  \frac{-1}{\sqrt{6}} \sum_{nlm}a_{nlm}(\tau) Q^{n}_{lm}(x^i), \\
\delta \phi (\tau, x^i) &= \frac{1}{\sqrt{6}}  \sum_{nlm} f_{nlm}(\tau) Q^{n}_{lm}(x^i). 
\end{align}
\end{subequations}
We denote $(n,\ell,m)$ collectively by $(n)$. The sums above begin with $n=2$ in a suitable choice of gauge. 

The differential equations for the complex extrema [cf. EI(A2)] allow the following behavior for their solutions near $\tau=0$:
\begin{equation}
\label{ptnrsp}
f_{(n)}(\tau) \sim A_{(n)} \tau^{-1\pm n}, \quad a_{(n)} \sim B_{(n)} \tau^{1\pm n} 
\end{equation}
where $ A_{(n)}$ is an arbitrary constant and $B_{(n)}$ is a constant linearly related to it [cf. EI(A3)].
The solutions with the minus signs are singular. The  field diverges at $\tau=0$ as does the perturbation in the geometry. Regularity corresponds to the plus signs. For these, the perturbations in both field and geometry vanish at the SP. (More precisely $a_{(n)}$ and $f_{(n)}$ vanish. Regularity may be expressed differently for different gauge invariant combinations of these variables.)  

{\cred The rate of growth of the fluctuations away from the bounce is defined by \eqref{ptnrsp} and \eqref{expnharmonics} given a quantum state that fixes the constant coefficients. This rate of growth determines the size of the region in the $\tau$-plane for which the fluctuations are small indicated roughly by the circle in Figure \ref{contour}.} 
We next examine the consequences of this for the Lorentzian perturbations.

\subsection{Lorentzian Fluctuations}

The behavior of the complex, perturbed saddle points along the broken contour $C_B(X)$ described in Section \ref{fuzzy}  and illustrated in  Figure \ref{contour} was investigated numerically in the Appendix of EI. The following generic behavior emerged for a mode $(n)$: The fluctuation vanishes at the SP. Then it oscillates {\crc in its ground state} along the contour while $|a(\tau)H_E(\tau)|<n$  where $H_E\equiv (1/a)(da/d\tau)$. This is the analog for saddle points of the Lorentzian condition that the wavelength of the fluctuation be inside the horizon. When the complex fluctuation leaves the horizon it { tends to a constant value}. The imaginary parts of  the $z_{(n)}(\tau)$ decay quickly in this period and the real part of the action of the saddle point $I^{(2)}$ tends to a constant. The fluctuations then satisfy the classicality condition that $S^{(2)}$ varies rapidly compared to  $I^{(2)}$.  The real values of the $z_{(n)}(\tau)$ become classical Lorentzian histories of fluctuations on the homogeneous and isotropic background. The probabilities of the fluctuations are $\exp[-2I^{(2)}(z_{(n)})/\hbar]$  to leading order in $\hbar$. Figure \ref{fluctuations} and the figures in the Appendix of EI illustrate all this explicitly. 

Thus, with an appropriate choice of contour, we have a unified picture of the complex fluctuation saddle points, the Lorentzian histories they predict, and the connection between them. This connection establishes the main result needed for the analysis in the text. 

Fluctuations vanish at the SP which is at the origin of the $\tau-$plane. They are therefore small in a region around the SP indicated by the circle in Figure \ref{contour}. Classical histories lie along approximately constant $x$ vertical lines with the scale factor $b$ increasing with increasing with $|y|$ along the curve. For bouncing universes $b$ reaches a minimum when the curve is closest to the SP. Fluctuations are therefore small at the bounce. Non-bouncing universes may have several singular regimes where $b\approx 0$. But only one of these can be near the SP at $y\approx 0$. The other must be at large $y$. Thus we justify the assumption made in  Section \ref{sah}: {\crh {\it The Lorentzian histories have one and only one range of three-surfaces where the fluctuations are small.   For bouncing universes this range is at the bounce. For non-bouncing universes this range is near one singularity.}}


\begin{thebibliography}{99}

\bibitem{Bol97} L.~Boltzmann, {\it Zu Hrn.~Zermelo's Abhandlung \"Uber
die mechanische Erkl\"arung Irreversibler Vorgange}, {\sl Ann.~Physik},
{\bf 60}, 392-398 (1897).

\bibitem{HH83}
J. B.~Hartle and S. W.~Hawking, \ttle{The Wave Function of the Universe,} {\sl Phys. Rev. D} {\bf 28}, 2960-2975 (1983).

\bibitem{GH07} M.~Gell-Mann and J.B.~Hartle, \ttle{\it Quasiclassical Coarse Graining and Thermodynamic Entropy}, {\sl Phys. Rev. A}, {\bf 76}, 022104 (2007) , quant-ph/0609190.

\bibitem{Har05} J.B.~Hartle, \ttle{\it The Physics of Now}, {\sl Am. J. Phys.}, {\bf 73}, 101-109
(2005), gr-qc/0403001.

\bibitem{Wei08} S.~Weinberg, {\it Cosmology}, (Oxford, OUP, 2008).

\bibitem{Bou11}  R.~Bousso, {\it Vacuum Structure and the Arrow of Time}, arXiv:1112.3341. 

\bibitem{HH11} T. Hertog and J.~Hartle, {\it Holographic No-Boundary Measure}, arXiv:1111.6090. 

\bibitem{Har09} J.B.~Hartle, {\sl The quasiclassical realms of this quantum universe}, arXiv:0806.3776.  A slightly shorter version is published in  {\sl Many Worlds?} edited by  S. Saunders, J.Barrett, A.Kent, and D. Wallace (Oxford University Press, Oxford, 2010). 

\bibitem{Haw84} S.W.~Hawking, \ttle{\it The quantum state of the universe},
{\sl Nucl.~Phys.~B}, {\bf 239}, 257-276 (1984).

\bibitem{HH85} J. J. Halliwell and S. W. Hawking,  \ttle{\it Origin of structure in the universe}, 
{\sl Phys. Rev. D}  {\bf 31}, 1777 (1985).

\bibitem{Page06} D.N.~Page, \ttle{\it Susskind's Challenge to the Hartle-Hawking No-Boundary Proposal and Possible Resolutions},  {\sl JCAP},  0701:004 (2007); arXiv:hep-th/0610199.

\bibitem{HLL93} S. W. Hawking, R. Laflamme, and G. W. Lyons, \ttle{\it Origin of time asymmetry}, {\sl  Phys. Rev. D} {\bf 47}, 5342-5356 (1993) (HLL). 


\bibitem{HHH08a} J.B.~Hartle, S.W. Hawking, and T. Hertog, \ttle{\it The no-boundary measure of the universe,} {\sl Phys. Rev. Lett.},  {\bf 100}, 202301 (2008), arXiv:0711.4630.

\bibitem{HHH08b} J.B.~Hartle, S.W. Hawking, and T. Hertog, \ttle{\it Classical universes of the no-boundary quantum state}, {\sl Phys. Rev. D} {\bf  77}, 123537 (2008), arXiv:0803.1663 (CU).

\bibitem{HHH10a} J.B.~Hartle, S.W. Hawking, and T. Hertog, \ttle{\it The No-Boundary Measure in the Regime of Eternal Inflation,} {\sl Phys. Rev. D}, {\bf 82}, 063510; arXiv:1001.0262 (EI).

\bibitem{HHH10b} J.B.~Hartle, S.W. Hawking, and T. Hertog, \ttle{\it Local Observation in Eternal Inflation}, {\sl Phys. Rev. Lett.}, {\bf 106}, 141302 (2011); arXiv:1009.2525.

\bibitem{Har95c} J.B.~Hartle, {\it Spacetime Quantum Mechanics and the
Quantum Mechanics of Spacetime} in {\sl 
Gravitation and Quantizations}, Proceedings of the 1992 Les Houches
Summer School, ed.~by B.~Julia and J.~Zinn-Justin, Les Houches Summer
School Proceedings Vol.~LVII, North Holland, Amsterdam (1995); 
gr-qc/9304006. A pr\'ecis of these lectures is given in gr-qc/9304006.

\bibitem{HHrules} J.B.~Hartle and T.~Hertog, {\sl Classical Prediction in Quantum Cosmology}, in preparation. 

\bibitem{HN64} F. Hoyle and J.V.~Narlikar, \ttle{\it Symmetric Electrodynamics and the Arrow of Time in Cosmology}, {\sl Proc. Roy. Soc. Lond.}, {\bf A277}, 1-23 (1964). 

\bibitem{CC04} S.~Carroll and J.~Chen, \ttle{\it Spontaneous Inflation and the Origin of the Arrow of Time},
arXiv:hep-th/0410270. 

\bibitem{KZ95} C.~Kiefer and H.D.~Zeh, \ttle{\it Arrow of Time in a Recollapsing Quantum Universe}, {\sl Phys. Rev. D}, {\bf 51},4145 (1995). 

\bibitem{Boj09} M.~Bojowald, {\it A Momentous Arrow of Time}, arXiv:0910:3200. 

\bibitem{Coc67} W.J.~Cocke, \ttle{\it Statistical Time Symmetry and Two-Time Boundary Conditions in Physics and Cosmology}, {\sl Phys.~Rev.} {\bf 160}, 1165 (1967).

\bibitem{GH93b} M.~Gell-Mann and J.B.~Hartle, {\it Time Symmetry and
Asymmetry in Quantum Mechanics and Quantum Cosmology}, in {\sl The Physical
Origins of Time Asymmetry}, ed. by J. Halliwell, J. P\'erez-Mercader, and W. Zurek, (CambridgeUniversity Press, Cambridge,1994); arXiv: gr-qc/9309012.

\bibitem{Pag84} D.~Page, {\it A fractal set of perpetually bouncing universes?}, {\sl Class. Quant. Grav.}, {\bf 1}, 417-427 (1984). For multiple bounces arising from quantum corrections see  V.T.~Gurovich and A.A.~Starobinsky, {\it Quantum Effects and Regular Cosmologies},  {\sl Zh. Eksp. Teor. Fiz.}, {\bf 77}, 1699, (1979) [
Eng. trans.:  {\sl Sov. Phys. JETP}, {\bf 50}, 844 (1979)].  

\bibitem{LB90} B.~Basu and D. Lynden-Bell, {\it A Survey of Entropy in the Universe},
{\sl QJRAS}, {\bf 31}, 359-369 (1990). 


\bibitem{ekpyrotic} J.~Khoury, B. A. Ovrut, P. J. Steinhardt, N. Turok , {\it Ekpyrotic universe: Colliding branes and the origin of the hot big bang}. Physical Review D {\bf 64}. 123522 (2001),arXiv:hep-th/0103239.


\bibitem{prebb} M.~Gasperini and G.~Veneziano, \ttle{\it Pre - big bang in string cosmology}, {\sl Astropart.\ Phys.\ } {\bf 1}  317 (1993), arXiv:hep-th/9211021; J.~Khoury, B.~A.~Ovrut, P.~J.~Steinhardt and N.~Turok, \ttle{\it The ekpyrotic universe: Colliding branes and the origin of the hot big bang}, {\sl Phys.\ Rev.\  D}, {\bf 64}, 123522, (2001), arXiv:hep-th/0103239; P.~J.~Steinhardt and N.~Turok, \ttle{\it Cosmic evolution in a cyclic universe}, {\sl Phys.\ Rev.\  D}, {\bf 65}, 126003 (2002), arXiv:hep-th/0111098.

\bibitem{Penrose} R. Penrose, {\it Cycles of Time: An Extraordinary New View of the Universe}, (Knopf, New York, 2011), V.G.~Gurzadyan and R. Penrose, {\it Concentric circles in WMAP data may provide evidence of violent pre-Big-Bang activity}, arXiv:1012:3706. 

\bibitem{SING} T.~Hertog and G.~T.~Horowitz,  \ttle{\it Towards a big crunch dual}, {\sl JHEP} {\bf 0407} 073 (2004),
arXiv:hep-th/0406134; \ttle{\it Holographic Description of AdS Cosmologies},  {\sl JHEP} {\bf 0504} 005 (2005),
arXiv:hep-th/0503071; B.~Craps, T.~Hertog and N.~Turok, \ttle{\it Quantum Resolution of Cosmological Singularities using AdS/CFT}, arXiv:0712.4180; J. ~L.~F.~Barbon and E.~Rabinovici,  \ttle{\it AdS Crunches, CFT Falls And Cosmological Complementarity}, arXiv:1102.3015.

\bibitem{HH90}  J.~Halliwell and J.B.~Hartle, \ttle{\it Integration contours for the no-boundary wave function of the universe}, {\it  Phys. Rev. D}, {\bf 41}, 1815 (1990).

\bibitem{Lyo92} G.W.~Lyons, {\it Complex solutions for the scalar field model of the universe}, {\sl Phys. Rev. D}, {\bf 46}, 1546-1550 (1992).


\end{thebibliography}
\end{document}